\documentclass[runningheads]{llncs}
\usepackage{xspace}
\usepackage{xcolor}
\usepackage{tikz}
\usepackage{pgfplots}
\usepackage{graphicx}
\usepackage{listings,chngcntr}
\usepackage[backref]{hyperref}
\usepackage{breakcites}
\usepackage{algorithm}
\usepackage{algorithmic}
\usepackage{wrapfig}
\usepackage{amssymb}
\usepackage{amsmath}
\usepackage{centernot}
\usepackage{makecell}
\usepackage[normalem]{ulem}

\usepackage[firstpage]{draftwatermark}
\usepackage[caption=false]{subfig}
\usepackage[english]{babel}
\usepackage[utf8]{inputenc}

\pgfplotsset{compat=1.6}
\usetikzlibrary{patterns,positioning,shapes.geometric}

\newcommand{\algref}[1]{Alg.~\ref{Alg:#1}}
\newcommand{\coderef}[1]{Listing~\ref{Cd:#1}}
\newcommand{\figref}[1]{Fig.~\ref{Fi:#1}}
\newcommand{\sectref}[1]{Section~\ref{Se:#1}}
\newcommand{\tableref}[1]{Tab.~\ref{Ta:#1}}
\newcommand{\algline}[1]{(Line~\ref{algte:#1})}
\newcommand{\alglines}[2]{(Line~\ref{algte:#1}--\ref{algte:#2})}
\newcommand{\codeline}[1]{Line~\ref{Line:#1}}

\newcommand{\highlightcode}[1]{\makebox[0pt][l]{\color{lightgray}\rule[-3pt]{#1\linewidth}{12pt}}}

\newcommand{\True}{\ensuremath{\mathit{true}}\xspace}
\newcommand{\False}{\ensuremath{\mathit{false}}\xspace}

\newcommand{\inangleb}[1]{ \ensuremath{\langle #1 \rangle}}

\newcommand{\app}[1]{\textsf{#1}\xspace}

\newcommand{\emp}[1]{\textsf{#1}}
\newcommand{\TT}[1]{{\tt #1}}
\newcommand{\EM}[1]{{\em #1}}

\newcommand{\tool}{\textsl{LLOR}\xspace}
\newcommand{\verifier}{\textsl{LLOV}\xspace}
\newcommand{\llvm}{\textsl{LLVM}\xspace}
\newcommand{\zth}{\textsl{Z3}\xspace}
\newcommand{\gpurepair}{\textsl{GPURepair}\xspace}
\newcommand{\sat}{\textsl{SAT}\xspace}
\newcommand{\mhs}{\textsl{mhs}\xspace}
\newcommand{\maxsat}{\textsl{MaxSAT}\xspace}

\newcommand{\toolname}{LLOR\xspace}

\newcommand{\llvmname}{LLVM\xspace}
\newcommand{\gpurepairname}{GPURepair\xspace}

\newcommand{\gmorcidlogo}{\href{https://orcid.org/0000-0002-3632-7801}{\protect\includegraphics[scale=0.4]{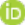}}}
\newcommand{\uborcidlogo}{\href{https://orcid.org/0000-0002-0076-1059}{\protect\includegraphics[scale=0.4]{images/orcid.pdf}}}
\newcommand{\sjorcidlogo}{\href{https://orcid.org/0000-0001-8070-1525}{\protect\includegraphics[scale=0.4]{images/orcid.pdf}}}
\newcommand{\ruorcidlogo}{\href{https://orcid.org/0000-0002-5290-3266}{\protect\includegraphics[scale=0.4]{images/orcid.pdf}}}

\begin{document}
\title{\toolname: Automated Repair of OpenMP Programs}

\author{Utpal Bora\inst{2}\textsuperscript{\uborcidlogo} \and
Saurabh Joshi\inst{3}\textsuperscript{\sjorcidlogo} \and
Gautam Muduganti\inst{1}\textsuperscript{\gmorcidlogo} \and
Ramakrishna Upadrasta\inst{1}\textsuperscript{\ruorcidlogo}
\thanks{The author names are in alphabetical order.}}

\authorrunning{Bora, Joshi, Muduganti et al.}

\institute{Indian Institute of Technology Hyderabad, India \\
\email {\{cs17resch01003,ramakrishna\}@iith.ac.in} \and
University of Cambridge, UK \\
\email{utpal.bora@cl.cam.ac.uk} \and
Supra Research \\
\email{sbjoshi@iith.ac.in}}

\SetWatermarkAngle{0}
\SetWatermarkText{\hspace*{5.5in}\raisebox{8.8in}{
    \href{https://doi.org/10.5281/zenodo.13886253}{\includegraphics{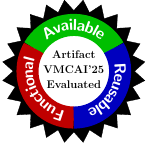}}
}}

\maketitle
\begin{abstract}
In this paper, we present a technique for repairing data race errors in parallel programs written in C/C++ and Fortran using the OpenMP API. Our technique can also remove barriers that are deemed unnecessary for correctness. We implement these ideas in our tool called \tool, which takes a language-independent approach to provide appropriate placements of synchronization constructs to avoid data races. To the best of our knowledge, \tool is the only tool that can repair parallel programs that use the OpenMP API. We showcase the capabilities of \tool by performing extensive experiments on $415$ parallel programs.

\keywords{OpenMP, Verification, Automated Repair, C, C++, Fortran}
\end{abstract}

\section{Introduction}
\label{Se:introduction}
Programs that can solve problems using multiple threads in parallel are defined as parallel programs. Parallelism can be achieved through concurrent computing using threads on CPUs or GPUs. There are two common approaches for achieving parallelism: Multiple Instruction, Multiple Data (MIMD) and Single Instruction, Multiple Data (SIMD). The OpenMP API~\cite{dagum1998openmp} provides a cross-platform abstraction for programs written in C/C++ and Fortran for both of these approaches.

A data race occurs in a parallel program when two or more threads are accessing the same memory location at the same time, and at least one of those accesses is a write. The behavior of such programs is unpredictable. Data races are one of the frequently encountered issues in concurrent computing. There is a substantial positive financial impact in identifying and repairing these errors early in the development cycle~\cite{boehm1988understanding}.

\coderef{race} without the highlighted line illustrates an OpenMP program that contains a data race. The data race exists because of the parallel reading and writing of the shared array \TT{data}. Placing a barrier (\TT{\#pragma omp barrier}) in the program at line \ref{codeline:race:barrier} mitigates the data race by ensuring that all the threads reach it before any of them can proceed further.

\coderef{race_for} presents another OpenMP program that has a similar data race. In this example, the data race is because of the \TT{parallel for} loop. OpenMP does not support barrier constructs inside a \TT{parallel for} loop. The data race can be avoided by adding the ordered clause to the \TT{for} loop (line \ref{codeline:race:ordered_for}) and placing the statements causing the data race inside an ordered region (\TT{\#pragma omp ordered}), as shown in line \ref{codeline:race:ordered}. Note that there can be multiple statements in this ordered region.

\begin{lstlisting}[caption=This OpenMP program without the highlighted line contains a data race, label={Cd:race}, captionpos=b, xleftmargin=2em, escapechar=\^]
int data[NUM_THREADS+1];
#pragma omp parallel {
    int id = omp_get_thread_num();
    int temp = data[id+1]; ^\label{Line:in:read}^
    ^\highlightcode{0.35}^#pragma omp barrier ^\label{codeline:race:barrier}^
    data[id] = temp; ^\label{Line:in:write}^
}
\end{lstlisting}

\begin{lstlisting}[caption=This OpenMP program without the highlighted snippets contains a data race inside the \TT{for} loop, label={Cd:race_for}, captionpos=b, xleftmargin=2em, escapechar=\^]
int data[count+1];
#pragma omp parallel for ^\highlightcode{0.13}^ordered^\label{codeline:race:ordered_for}^
for (int i=0; i<count; i++) {
    int temp = data[i+1]; ^\label{Line:in:read2}^
    ^\highlightcode{0.35}^#pragma omp ordered ^\label{codeline:race:ordered}^
    data[i] = temp; ^\label{Line:in:write2}^
}
\end{lstlisting}

In this tool paper, we make the following contributions:

\begin{itemize}
\item We introduce our tool, \tool, which takes a language-independent approach to automatically fix data race errors in programs written using the OpenMP API.
\item For OpenMP programs that contain \TT{parallel} regions, our paper proposes a technique to provide barrier placements to avoid data races.
\item For OpenMP programs that contain \TT{parallel for} loops, our paper proposes a technique to identify the statements that need to be placed in an ordered region to avoid data races.
\item In addition to the above two techniques that introduce synchronization constructs to avoid data races, our paper proposes a methodology to remove existing barriers and ordered regions inserted by the programmer if deemed unnecessary.
\item Our paper showcases the practical differences between the two different solver strategies (\mhs and \maxsat) that are employed during the repair process.
\item We showcase the effectiveness of our tool by performing extensive experimental evaluation on $415$ parallel programs that use the OpenMP API. This benchmark set consists of $235$ C/C++ programs and $180$ Fortran programs, highlighting the language versatility of \tool.
\end{itemize}

To the best of our knowledge, ours is the only technique and tool that can propose a fix for parallel programs written using the OpenMP API. In \sectref{llor}, we describe the working of \tool in fine detail, discuss related work in \sectref{relatedwork}, and in \sectref{experiments}, we provide a detailed background on the experimental setup and present the results summary of running \tool against the benchmark suite.

\section{\toolname}
\label{Se:llor}

\subsection{\toolname architecture and workflow}
\label{Se:workflow}
\begin{figure}[htp]
\centering
\includegraphics[scale=0.25]{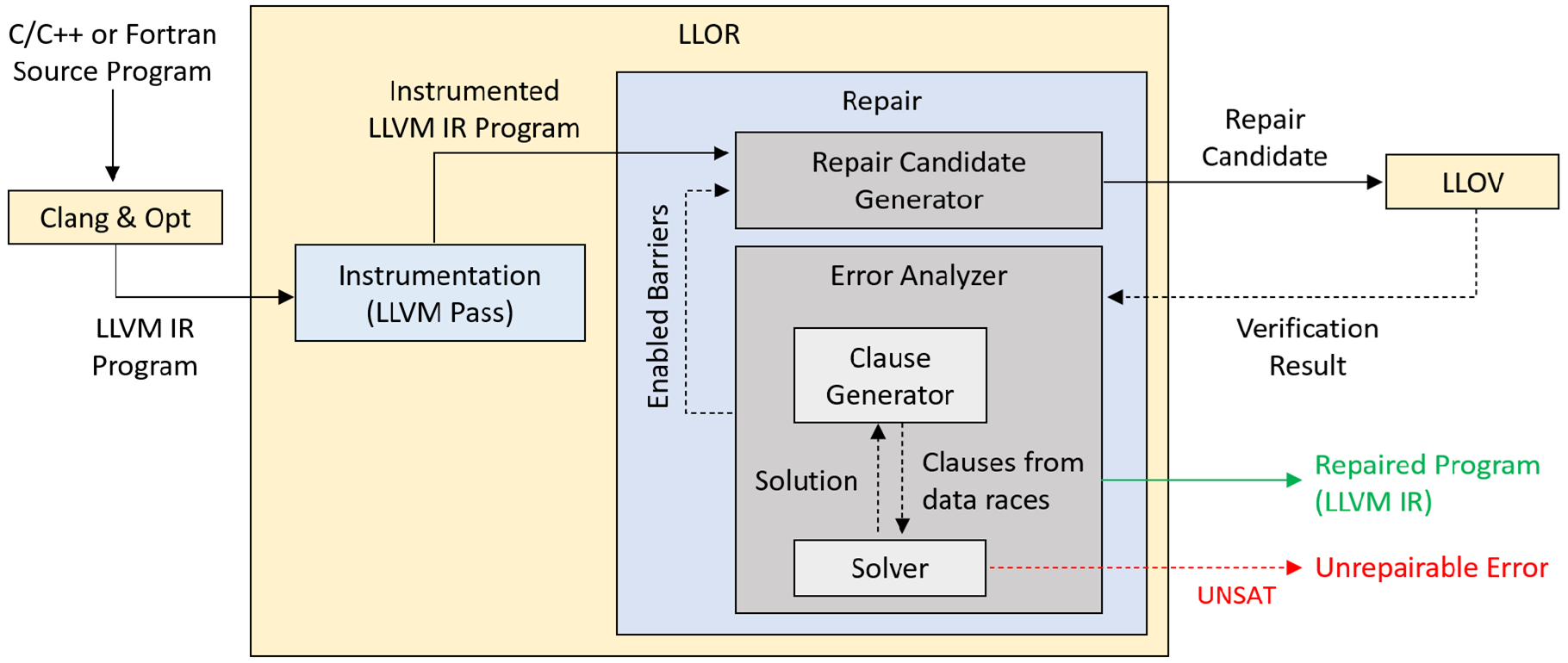}
\caption{Architecture of \tool showcasing the various components involved in the repair process. The solid lines represent the source code, and the dashed lines represent the information flow between the components.}
\label{Fi:repairer_workflow}
\end{figure}

The implementation of \tool leverages \verifier~\cite{bora2020llov}, a state-of-the-art data race checker, as depicted in \figref{repairer_workflow}. \verifier identifies data races in C/C++ and Fortran programs. It is built as an analysis pass using the \llvm compiler infrastructure. The architecture of \tool consists of two components: Instrumentation and Repair.

The instrumentation component adds metadata that indicates the possible locations of barriers in the case of \TT{parallel} regions. In the case of \TT{parallel for} loops, this metadata marks the possible statements that have to be a part of an ordered region.

The repair component takes an iterative approach. In each iteration, \verifier is called with a repair candidate to check if the program is error-free. If \verifier identifies any data races, constraints are generated from these errors to avoid them in subsequent iterations. The \EM{Solver} is called with these constraints to obtain a solution that determines which synchronization constructs have to be enabled or disabled while generating the next repair candidate.

If \tool can find an error-free repair candidate, it generates the \llvmname intermediate representation (\llvmname IR)~\cite{lattner2004llvm} of the repaired program and a summary file. The summary file contains the necessary changes to repair the program, along with the source location details of the original C/C++ or Fortran input program. In this paper, we have used \verifier~\cite{bora2020llov} as the verifier, but in principle, any OpenMP verifier~\cite{basupalli2011ompverify,ye2018using,chatarasi2016static,atzeni2018sword,eichenberger2013ompt,valgrind2007helgrind} can be used for this technique to work.

\subsection{Instrumentation}
\label{Se:instrumentation}
Since \tool attempts to fix errors caused only due to data races, it proposes a solution that involves either adding new barriers in a \TT{parallel} region or creating an ordered region for a subset of instructions inside a \TT{parallel for} loop. It also tries to remove unnecessary barriers and ordered regions. Data races can only be caused when multiple threads read and write the same shared variable. Taking this into account, the instrumentation component identifies all the instructions that are either reading or writing a shared variable.

If these instructions are in a \TT{parallel} region, \tool generates possible repair candidates by inserting barriers in front of these instructions. If these instructions are in a \TT{parallel for} loop, \tool generates possible repair candidates by selecting the smallest subset of instructions that need to be placed inside an ordered region. Existing barriers and ordered regions in the program are removed to see if they are indeed required for program correctness. Using this technique significantly reduces the search space of repair candidates since the number of instructions that involve a shared variable is typically much fewer than the total number of instructions in the program.

Consider the OpenMP program presented in \coderef{race}. The instrumentation component identifies that the shared variable \TT{data} is being accessed at \codeline{in:read} and \codeline{in:write} and adds metadata indicating that a barrier might be necessary before these instructions. Similarly, in \coderef{race_for}, it identifies that the shared variable \TT{data} is being accessed at \codeline{in:read2} and \codeline{in:write2} and that these instructions are inside a for loop. Metadata indicating that these instructions might have to be placed inside an ordered region is added in this case. Each of these instructions is marked with a Boolean variable that indicates whether the barrier should be enabled or not in the case of \TT{parallel} regions. In the case of \TT{parallel for} loops, these variables indicate if the instruction should be placed inside an ordered region. These variables are referred to as \EM{barrier variables}. By default, these barrier variables are set to \TT{false} to check if the program can be concluded as error-free without introducing any additional synchronization constructs.

\subsection{The Repair Algorithm}
\label{Se:repair_algorithm}
The repair technique used in \tool is depicted in \algref{llor}. The input to this algorithm is the \llvmname IR program obtained after instrumentation. We walk through this algorithm using the example presented in \coderef{algo_barrier}. Without the highlighted portions, this program contains two data races: one caused by memory accesses of \TT{data\_a} at \codeline{algob:asource} and \codeline{algob:asink} and the other caused by memory accesses of \TT{data\_b} at \codeline{algob:bsource} and \codeline{algob:bsink}. The instrumentation component identifies that these four statements either read or write a shared variable. For each statement, a barrier variable is introduced that indicates whether a synchronization construct is needed before the statement.

\begin{lstlisting}[caption=Instrumented OpenMP program that contains multiple data races in a \TT{parallel} region. The highlighted parts represent the changes in the second iteration., label={Cd:algo_barrier}, captionpos=b, xleftmargin=2em, escapechar=\^]
int data_a[NUM_THREADS+1];
int data_b[NUM_THREADS+1];
#pragma omp parallel {
    int id = omp_get_thread_num();
    int temp_a = data_a[id+1];  // b1=false ^\label{Line:algob:asource}^
    int temp_b = data_b[id+1];  // b2=false ^\label{Line:algob:bsource}^
    ^\highlightcode{0.35}^#pragma omp barrier
    data_a[id] = temp_a;        // b3=false ^\highlightcode{0.11}^(true) ^\label{Line:algob:asink}^
    data_b[id] = temp_b;        // b4=false ^\label{Line:algob:bsink}^
}
\end{lstlisting}

\begin{lstlisting}[caption=Instrumented OpenMP program that contains multiple data races in a \TT{parallel for} loop. The highlighted parts represent the changes in the second iteration., label={Cd:algo_ordered}, captionpos=b, xleftmargin=2em, escapechar=\^]
int data_a[count+1];
int data_b[count+1];
#pragma omp parallel for ^\highlightcode{0.13}^ordered
for (int i=0; i<count; i++) {
    int temp_a = data_a[i+1];  // b1=false
    int temp_b = data_b[i+1];  // b2=false
    ^\highlightcode{0.37}^#pragma omp ordered {
        data_a[i] = temp_a;    // b3=false ^\highlightcode{0.11}^(true)
        data_b[i] = temp_b;    // b4=false
    }
}
\end{lstlisting}

The algorithm iteratively calls the verifier \alglines{startloop}{endloop} until it either finds a solution or concludes that the program cannot be repaired. In every iteration, $\mathit{GenerateRepairCandidate}$ \algline{repaircandidate} generates a repair candidate using the error traces seen till then. The first iteration of this algorithm always verifies the input program without enabling any synchronization constructs since all barrier variables are set to \TT{false} by default. The verifier returns the verification $\mathit{result}$ along with an error trace $\pi$ if it identifies an error in the program. If the verifier is unable to find an error \algline{repaired} with the proposed solution $\mathit{sol}$, then the algorithm exits the loop, and the instrumented LLVM IR program constrained with $\mathit{sol}$ is returned \algline{returnsol}. If the verifier returns an error that is not a data race \algline{asserterror}, then \algref{llor} terminates with an error stating that it cannot repair the program.

The error trace provided by the verifier contains the line numbers of the statements involved in the data race. Introducing a synchronization construct between these lines will mitigate the data race. However, there could be multiple locations between these line numbers where the synchronization construct could be introduced. The algorithm aims to identify the optimum placement of these synchronization constructs since there are performance penalties for every synchronization construct that is introduced.

$\mathit{GenerateClause}$ \algline{generateclause} generates a \EM{positive monotone clause} (a clause having only positive literals) for every data race error using the disabled barrier variables that exist between the line numbers obtained from the error trace. These clauses are added to the constraint $\varphi$ \algline{addclause}. If $\varphi$ is satisfiable by $\mathit{Solve}$ \algline{solve}, we proceed to generate the repair candidate. Since $\varphi$ consists of only positive monotone clauses, it will always be satisfiable as such a formula can be satisfied by an assignment that sets all the literals to true. The only exception is the case when we encounter an empty clause, which is generated when the data race occurs due to a write-write conflict on the same line. In these cases \algline{unsat}, the algorithm terminates, stating that the program cannot be repaired.

\begin{algorithm}[t]
\caption{The Repair Algorithm}
\label{Alg:llor}

\begin{algorithmic}[1]
\STATE \emp{Input:} Instrumented Program $P$
\STATE \emp{Output:} Repaired Program $P_{sol}$

\STATE $\mathit{\varphi := true}$ \label{algte:constraint}

\LOOP \label{algte:startloop}
\STATE $\mathit{\inangleb{res,sol} := Solve(\varphi)}$ \label{algte:solve}
\IF {$\mathit{res = UNSAT}$} \label{algte:unsat}
\PRINT \emp{Error:} Program cannot be repaired \label{algte:unsaterror}
\RETURN \emp{errorcode}
\ENDIF
\STATE $\mathit{P_{sol} := GenerateRepairCandidate(P,sol)}$  \label{algte:repaircandidate}
\STATE $\mathit{\inangleb{result, \pi} := Verify(P_{sol})}$ \label{algte:verify}

\IF {$\mathit{result = SAFE}$} \label{algte:repaired}
\STATE \emp{break}
\ENDIF

\IF {$\mathit{result \neq RACE}$} \label{algte:asserterror}
\PRINT \emp{Error:} Program cannot be repaired
\RETURN \emp{errorcode}
\ENDIF

\STATE $\mathit{c := GenerateClause(\pi)}$ \label{algte:generateclause}
\STATE $\mathit{\varphi := \varphi \cup \{c\}}$ \label{algte:addclause}
\ENDLOOP \label{algte:endloop}

\RETURN $P_{sol}$ \label{algte:returnsol}
\end{algorithmic}

\end{algorithm}

Based on the error trace information obtained from the verifier for \coderef{algo_barrier}, two clauses are generated: $b_2 \vee b_3$ and $b_3 \vee b_4$. There are multiple solutions available for $\mathit{Solve}$ to make these clauses satisfiable. Let us consider that $\mathit{Solve}$ chooses the optimum solution, which is to enable $b_3$ and disable the rest. The highlighted portions of \coderef{algo_barrier} are then introduced to obtain the repair candidate for the next iteration. The verifier does not identify any data race errors with this repair candidate, so the repair algorithm terminates successfully by returning the repaired program.

In the case of \TT{parallel for} loops, the repair candidate generator does not introduce barriers. Instead, it creates an ordered region that starts with the earliest statement associated with an enabled barrier variable to the last statement that uses a shared variable. Note that there can be only one ordered region per loop. Because of this restriction, our technique may include statements that do not even use a shared variable in the ordered region. \coderef{algo_ordered} shows an example of how the repair algorithm works for \TT{parallel for} loops.

There are several ways to obtain $\mathit{sol}$ from $\varphi$ in $\mathit{Solve}$ \algline{solve}. A basic implementation of $\mathit{Solve}$ could involve using a \sat solver. However, using a \sat solver could cause the repair algorithm to enable several unnecessary synchronization constructs since the solver does not guarantee optimality. Instead, we propose two different strategies for implementing $\mathit{Solve}$. The \maxsat strategy implements this as a partial \maxsat problem~\cite{fu2006solving} with $\varphi$ as hard clauses and $\{\neg b_1,\dots,\neg b_m\}$ as soft clauses. The \mhs strategy computes a minimal-hitting-set (\mhs) over $\varphi$ using a polynomial time greedy algorithm~\cite{johnson1974approximation}.

\subsection{Properties of our algorithm}
\label{Apx:properties}

\subsubsection{Soundness}
Assuming that the verifier used in $\mathit{Verify}$ is sound and the solver used in $\mathit{Solve}$ is sound and complete, if \algref{llor} returns a repaired program $P_{sol}$, $P_{sol}$ would not have any data races. This is evident because a sound verifier \algline{verify} would declare $P_{sol}$ as $\mathit{SAFE}$ \algline{repaired} only if the program indeed does not have any errors under the constraints provided by $\mathit{sol}$.

\subsubsection{Termination}
The number of barrier variables in the instrumented program $P$ is finite. $\varphi$ contains all the clauses generated from the error traces seen until then, and $\mathit{sol} \implies \varphi$ since all the clauses represented by $\varphi$ are satisfied by $\mathit{sol}$. Every clause generated \algline{generateclause} by \algref{llor} is such that $\varphi \centernot \implies c$ because $\mathit{Verify}$ \algline{verify} will never result in an error trace that has already been encountered before. Given that the number of barrier variables is finite and error traces do not repeat, the number of clauses that can be generated is also finite, ensuring that the algorithm terminates. Because the barrier variables correspond to program text, a barrier variable set to \True (resp. \False) would enable (resp. disable) the synchronization construct for all iterations of a loop.

\subsubsection{Completeness}
Let $P$ be the input program such that it only has data race errors that can be fixed by the introduction of synchronization constructs discussed in \sectref{introduction}. If the solver used in $\mathit{Solve}$ and the verifier used in $\mathit{Verify}$ are sound and complete, \algref{llor} would always find a repaired program $P_{sol}$.

\algref{llor} exits when it either encounters an error that is not a data race \algline{asserterror}, a repaired program has been obtained \algline{repaired}, or an empty clause has been encountered \algline{unsat}. Since $P$ contains only data races that can be avoided, \algref{llor} would terminate only when a solution has been obtained \algline{repaired}. Given that the algorithm is sound, the repaired program $P_{sol}$ obtained with $\mathit{sol}$ as the solution is guaranteed to be free of data races.

\subsubsection{Optimality}
When the \maxsat strategy is used in the implementation of $\mathit{Solve}$ \algline{solve}, the solution obtained ($\mathit{sol}$) is guaranteed to be optimum by the definition of the \maxsat algorithm \cite{fu2006solving,marques2007using}.

When the \mhs strategy is used, the solver need not enable the least number of synchronization constructs. After the repaired program is obtained, the \maxsat solver is invoked once with the clauses generated ($\varphi$) to ensure that the solution obtained is indeed optimum. \\

We implement this algorithm in our tool \tool, which uses \verifier as the underlying verifier. The technique behind \verifier is sound but not complete. However, \verifier is not sound because of programs like \coderef{llov_fortran}. Since \verifier is neither sound nor complete, \tool cannot guarantee soundness or completeness.

\section{Related Work}
\label{Se:relatedwork}

\subsection{Verification of Parallel Programs}
\label{Se:rw_verification}

Identifying data races in parallel programs has been an active area of research for quite a while. Some of the early techniques proposed in this area were based on the \EM{happens-before} relation defined by Lamport~\cite{lamport2019time}. Mellor-Crummey~\cite{mellor1991fly} introduced a new protocol that uses Offset-Span Labeling of the nodes in a fork-join graph that represents the concurrency relationships among threads in an execution of a fork-join program.

\app{Eraser}~\cite{savage1997eraser} tries to address the limitations of the tools built using the happens-before relation by introducing a new Lockset algorithm. In addition to these, several other techniques~\cite{dinning1991detecting,netzer1991race,perkovic1996online,joshi2012new} have been introduced to detect races in parallel programs. \app{Eraser}~\cite{savage1997eraser}, \app{Relay}~\cite{voung2007relay}, \app{Locksmith}~\cite{pratikakis2011locksmith}, and \app{RacerX}~\cite{engler2003racerx} use the Lockset algorithm to detect races in the pthreads execution model of C/C++ programs. \app{RacerD}~\cite{blackshear2018racerd} is a static analysis tool for Java programs. Tools like \app{GPUVerify}~\cite{betts2012gpuverify,betts2015design}, \app{ESBMC-GPU}~\cite{monteiro2018esbmc}, \app{VerCors}~\cite{blom2014specification,amighi2015specification}, \app{PUG}~\cite{li2010scalable}, and \app{GKLEE}~\cite{li2012gklee} propose techniques to detect data races in GPU programs.

Tools like \app{ompVerify}~\cite{basupalli2011ompverify}, \app{DRACO}~\cite{ye2018using}, \app{PolyOMP}~\cite{chatarasi2016static,chatarasi2017extended}, \app{SWORD}~\cite{atzeni2018sword}, \app{OMPT}~\cite{eichenberger2013ompt}, \app{Helgrind}~\cite{valgrind2007helgrind}, and \app{LLOV}~\cite{bora2020llov} have been introduced to identify data races in programs using the OpenMP API. Some of these tools~\cite{basupalli2011ompverify,ye2018using,chatarasi2016static,chatarasi2017extended,bora2020llov} use only static techniques, while some~\cite{atzeni2018sword,eichenberger2013ompt,valgrind2007helgrind} employ both static and dynamic techniques. \verifier is a static data race detection tool built on top of the \llvm~\cite{lattner2004llvm} compiler infrastructure, which uses \EM{RDG (Reduced Dependence Graph)} to model parallel regions of an \llvmname IR program and infer race conditions based on the presence of data dependencies.

\subsection{Automatic Program Repair}
\label{Se:rw_programrepair}

Using automatic techniques to repair bugs in programs has been another area of research that aligns closely with our work. The research works in this area can be broadly classified into techniques that attempt to repair sequential programs and techniques that attempt to repair concurrent programs.

The approaches that focus on sequential programs~\cite{jobstmann2005program,griesmayer2006repair,chandra2011angelic,malik2011constraint} take a set of assertions that need to be met and an input program that fails one or more of these assertions. The technique then modifies a small subset of statements in the input program to ensure that all the provided assertions pass.

The approaches that focus on concurrent programs~\cite{vechev2010abstraction,muzahid2010atomtracker,deshmukh2010logical,vcerny2011quantitative,jin2011automated,joshi2014automatically,joshi2015property,joshi2021gpurepair} work on an input program that passes all the assertions when run sequentially but has inconsistent behavior when run in parallel using multiple threads. This inconsistent behavior is because of the interleaving of threads. Repairing such programs involves introducing critical regions, locks, and problem-specific synchronization constructs.

In this paper, we instrument the given concurrent input program with the possible locations of the synchronization constructs and generate repair candidates based on the error traces obtained from the verifier. Similar techniques~\cite{joshi2014automatically,joshi2015property,joshi2021gpurepair} have been used in the past to repair concurrent programs. To the best of our knowledge, \tool is the only tool that can fix data races in OpenMP programs written in C/C++ and Fortran.

\subsection{Comparison with \gpurepairname}
\label{Se:comparison}

The counter-example driven repair technique used in this paper is similar to one of our earlier works~\cite{joshi2021gpurepair} that focuses on the automated repair of GPU kernels and the implementation of the technique in our tool, \gpurepair. However, the motivations behind \gpurepair and \tool are fundamentally different, causing the repair algorithm to differ in a non-trivial fashion.

GPU kernels work on the principle of Single Instruction Multiple Data (SIMD). The threads created in the kernel are organized as blocks. Blocks consist of warps. The threads within a warp generally execute in a lock-step manner. There could be data races within the threads of a block. Data races in GPU kernels are avoided using barrier constructs. However, there might be a scenario where some of the threads in a block reach the barrier, and some do not, resulting in a deadlock. This problem is called barrier divergence. \gpurepair aims to fix data race and barrier divergence errors in CUDA and OpenCL kernels.

\tool aims at fixing data races in OpenMP programs that have \TT{omp parallel} regions and \TT{omp parallel for} loops. Since OpenMP programs with these constructs work on the principle of Multiple Instruction Multiple Data (MIMD), the problem of barrier divergence does not exist here. Besides proposing barrier placements for \TT{omp parallel} regions, our technique also identifies the statements that need to be placed within ordered regions for \TT{omp parallel for} loops.

Since \gpurepair has to repair errors due to data race and barrier divergence errors, the repair algorithm has to evaluate which barriers need to be removed and which barriers need to be added. The algorithm achieves this by generating positive monotone clauses for data race errors and negative monotone clauses for barrier divergence errors. However, in the case of \tool, the repair algorithm generates only positive monotone clauses.

The instrumentation component of \tool instruments the LLVM IR generated from the OpenMP programs. On the other hand, \gpurepair uses Boogie~\cite{barnett2005boogie} as an intermediate representation which is generated from GPU kernels. \gpurepair adds the instrumentation also in Boogie. These differences give rise to an entirely different implementation stack, with \tool using C++ while \gpurepair uses C\#. Another notable difference in \tool is that we do not add conditional barriers but instead add metadata on possible locations of barriers.

\section{Experiments}
\label{Se:experiments}
In this section, we present the experimental results of running \tool on various C/C++ and Fortran benchmarks. The source code of \tool is available at \cite{llorsrc}. The artifacts used to reproduce the results of this paper are available at \cite{vmcaiartifact}.

\subsection{Experimental Setup}
\label{Se:experimental_setup}
Several tools are involved in the pipeline of \tool, as introduced in \sectref{workflow}. The instrumentation component is developed as an LLVM pass and is therefore built in C++. The repair component is outside the LLVM infrastructure and is built using the .NET Framework with C\# as the programming language. We use the \zth solver~\cite{de2008z3,bjorner2015nuz} to solve the constraints generated during the repair phase. The tools used in \tool and their versions are:
 \href{https://github.com/llvm/llvm-project/tree/release/12.x/llvm}{\app{LLVM} 12.0.0},
 \href{https://github.com/llvm/llvm-project/tree/release/12.x/clang}{\app{Clang} 12.0.0},
 \href{https://github.com/flang-compiler/flang}{\app{Flang} 12.0.1}, and
 \href{https://github.com/utpalbora/LLOV/tree/llov-v0.3}{\verifier 0.3}.

The experiments were performed on \emp{Standard\_F2s\_v2 Azure\textsuperscript{\textregistered} virtual machine}, which has 2 vCPUs and 4 GiB of memory. More details on the virtual machine can be found at \cite{azuref2sv2}. A total of $415$ programs ($235$ C/C++ and $180$ Fortran) were considered for the evaluation of \tool. This benchmark set consists of programs from the DataRaceBench~\cite{liao2017dataracebench} test suite, Exascale~\cite{exascale} project, Rodinia~\cite{che2009rodinia} test suite, and Parallel Research Kernels~\cite{parres}, along with additional benchmarks introduced while developing \tool. \tableref{benchmark_summary} summarizes the distribution of the benchmark set.

The benchmarks from DataRaceBench and \tool test suites have one source file per benchmark, and the remaining have multiple source files per benchmark. For benchmarks that have a single source file, the experiments were performed with a timeout of $300$ seconds for each benchmark. For benchmarks that have multiple source files, the experiments were performed with a timeout of $60$ seconds for each source file and an overall timeout of $1800$ seconds for the benchmark. Each benchmark has been executed $3$ times, and the average time of these $3$ runs is taken into consideration. We used the average since there was a negligible difference between the median and the average.

\begin{table}[t]
\caption{Benchmark Summary}
\label{Ta:benchmark_summary}
\begin{center}

\def\arraystretch{1.1}
\setlength\tabcolsep{7pt}

\begin{tabular}{|l|r|r|r|}
\hline

\multicolumn{1}{|c|}{\textbf{Source}} & \multicolumn{3}{|c|}{\textbf{Programs}} \\ \hline
\multicolumn{1}{|c|}{\textbf{}} & \multicolumn{1}{|c|}{\textbf{C/C++}} & \multicolumn{1}{|c|}{\textbf{Fortran}} & \multicolumn{1}{|c|}{\textbf{Total}} \\ \hline \hline
DataRaceBench & $181$ & $168$ & $349$ \\ \hline
Exascale Project & $8$ & $0$ & $8$ \\ \hline
Rodinia & $18$ & $0$ & $18$ \\ \hline
Parallel Research Kernels & $11$ & $0$ & $11$ \\ \hline
Other Large Benchmarks & $5$ & $0$ & $5$ \\ \hline
\toolname Test Suite & $12$ & $12$ & $24$ \\ \hline

\end{tabular}
\vspace{-20pt}

\end{center}
\end{table}

\subsection{Results}
\label{Se:results}

The results obtained from running \tool against the benchmark suite are summarized in \tableref{results}. The table categorizes the results into three categories based on the output of \verifier. The \hyperlink{racefree}{first category} includes all the programs for which \verifier concluded that there were no errors. On manual inspection, we found out that for some of the Fortran programs (e.g., \path{baseline_fortran/B01_simple_race.f95}, \path{dataracebench_fortran/DRB002-antidep1-var-yes.f95}), \verifier was not detecting a data race, even if one existed. Since the soundness and completeness guarantees of \tool are modulo the completeness of the verifier, \tool could not propose a fix for these programs. On the other hand, because of this behavior, \tool suggests removing existing barriers or ordered regions in some Fortran programs even though they are necessary. Note that this behavior of \verifier is limited to Fortran programs, and we did not come across any C/C++ programs that were impacted by this.

For \hyperlink{racefree_changes}{$11$ programs} in this category, \tool recommended changes. $3$ of these were Fortran programs (\path{baseline_fortran/B05_incorrect_barrier.f95}, \\ \path{baseline_fortran/B07_racefree.f95}, \\ \path{dataracebench_fortran/DRB110-ordered-orig-no.f95}) that \verifier had incorrectly declared as race-free when the barrier or ordered region was removed. For the remaining $8$ programs, \tool was correct in suggesting the removal of barriers and ordered regions.

The \hyperlink{racedetected}{second category} includes the programs for which \verifier had identified data races. Out of the $147$ programs in this category, \tool was able to fix \hyperlink{repaired}{$107$ programs} correctly. \tool was unable to repair \hyperlink{repairerror}{$16$ programs} since it could not find any assignment that satisfied the clauses generated during the repair phase. $5$ of these were C/C++ programs, and $11$ were Fortran programs. Upon manual inspection, we found that $11$ of these were similar, where an integer value was incremented through a static pointer variable. In the LLVM IR code, the address of the static pointer was copied to local variables, and the value was updated through these local variables. The instrumentation component of \tool could not track these operations, causing it to not instrument instructions that should have been instrumented. $2$ programs could not be fixed since multiple threads were writing different values to the same shared variable in parallel. No placement of barriers can fix such programs. The remaining $3$ programs had complex OpenMP constructs like \EM{teams} and \EM{target}, because of which simple barrier placements could not fix the data races. \hyperlink{unsupported}{$15$ programs} were using OpenMP constructs like \EM{sections} and \EM{simd}. These programs are beyond the scope of \tool since barriers and ordered regions cannot fix data races in these programs. \tool clearly states that these programs are unsupported when provided as input. \hyperlink{timedout}{$9$ programs} timed out.

\begin{table}[t]
\caption{Count of programs grouped by category}
\label{Ta:results}
\begin{center}

\def\arraystretch{1.1}
\setlength\tabcolsep{7pt}

\begin{tabular}{|l|r|r|}
\hline

\multicolumn{1}{|c|}{\textbf{Source}} & \multicolumn{1}{|c|}{\textbf{C/C++}} & \multicolumn{1}{|c|}{\textbf{Fortran}} \\ \hline
\multicolumn{1}{|c|}{Total Benchmarks} & $235$ & $180$ \hypertarget{racefree} \\ \hline  \hline
I. No data races identified by \verifier & $75$ & $129$ \\ \hline
\quad No changes made by \tool & $71$ & $122$ \hypertarget{racefree_changes} \\
\quad Changes recommended by \tool & $4$ & $7$ \hypertarget{racedetected} \\ \hline \hline
II. Data races identified by \verifier & $117$ & $30$ \hypertarget{repaired} \\ \hline
\quad Repaired by \tool & $92$ & $15$ \hypertarget{repairerror} \\
\quad Could not be repaired by \tool & $5$ & $11$ \hypertarget{timedout} \\
\quad Timeouts & $9$ & $0$ \hypertarget{unsupported} \\ 
\quad Unsupported & $11$ & $4$ \hypertarget{llovunsupported} \\ 
\hline \hline
III. Unsupported by \verifier & $43$ & $21$ \hypertarget{llorunsupported} \\ \hline
\quad Unsupported by \tool & $37$ & $11$ \hypertarget{compileerror} \\ 
\quad Compilation errors & $0$ & $10$ \hypertarget{verifyerror} \\
\quad Verification errors & $6$ & $0$ \\\hline

\end{tabular}
\vspace{-20pt}

\end{center}
\end{table}

The \hyperlink{llovunsupported}{final category} includes the programs that are either unsupported by \verifier or failed compilation. There are $64$ programs in this category. The \app{flang} version that we used for experimentation threw compilation errors for \hyperlink{compileerror}{$10$ Fortran programs} and did not generate \llvm IR. Because of that, we could not attempt to repair them. \verifier does not support the verification of OpenMP programs that use task-based parallelism. Since \tool depends on \verifier in the repair process, it could not generate any repair candidates for \hyperlink{llorunsupported}{$48$ programs}. \verifier either timed out or threw a runtime error for \hyperlink{verifyerror}{$6$ programs}.

\subsection{Solver Comparison}
\label{Se:solver_comparison}

As mentioned in \sectref{repair_algorithm}, \tool offers two different ways of solving the clauses generated from the error traces. The default approach uses the minimal-hitting-set (\mhs) strategy. The user can switch to the \maxsat strategy if they wish to. The \maxsat strategy is computationally heavier than the \mhs strategy since the \mhs solver uses a polynomial time algorithm, whereas the \maxsat solver makes multiple queries to the SAT solver. In most practical cases, the \mhs strategy would provide a similar solution to the \maxsat strategy. Because of this, the \mhs strategy has a performance advantage over the \maxsat strategy.

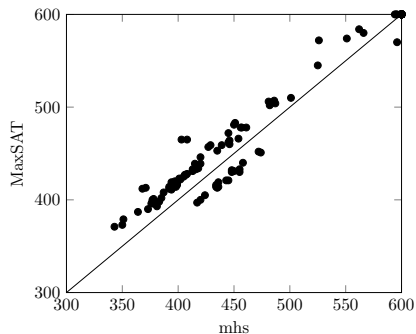
\begin{wrapfigure}{rt}{0.45\textwidth}
\centering

\begin{tikzpicture}[scale=0.65]
    
\pgfplotstableread{figures/data/time_mhs_maxsat.dat}{\data}
\selectcolormodel{gray}

\begin{axis}[
    xlabel={mhs},
    ylabel={MaxSAT},
    scaled ticks = false,
    tick label style={
        /pgf/number format/fixed,
        /pgf/number format/precision=3
    },
    xmin=300,
    xmax=600,
    ymin=300,
    ymax=600
]

\addplot[
    scatter=true,
    only marks,
    mark=*,
    scatter src=explicit symbolic,
    scatter/classes={
        a={mark=*}
    }
]
table[x=mhs,y=maxsat,meta=label]{\data};
\addplot [black,samples at={0,1}] {x};
\draw [black,solid] (rel axis cs:0,0) -- (rel axis cs:1,1);

\end{axis}
\end{tikzpicture}
\vspace{-15pt}
\caption[mhs vs. MaxSAT; Runtime in milliseconds]
    {\tabular[t]{@{}l@{}}mhs vs. MaxSAT \\ Runtime in milliseconds\endtabular}
\label{Fi:solver_time}
\vspace{-15pt}

\end{wrapfigure}

The behavior of the solver also impacts which strategy performs better. Consider the clause $a \vee b$. The \mhs solver could choose $a$ to be the solution, and the \maxsat solver could choose $b$ to be the solution. Both of the solutions are valid for the clause, but choosing $b$ could fix the program, and choosing $a$ may not, thus forcing more iterations. Because of these reasons, we notice that for some benchmarks, the \mhs strategy performs better, and the \maxsat strategy performs better for others.

Out of the $415$ benchmarks, only $133$ benchmarks did not result in a timeout and needed the use of a solver strategy (either \mhs or \maxsat). In $93$ of these benchmarks, the \mhs strategy performed better than the \maxsat strategy. \figref{solver_time} shows the comparison of these benchmarks. Most of the benchmarks executed within a second, so \figref{solver_time} is limited to $600$ milliseconds on the x-axis and y-axis to showcase those benchmarks. Any benchmark that took more than $600$ milliseconds has been plotted on the top-right corner of the graph.

Additional experiments and analysis are included in \ref{Apx:results}, which provides more experimental details like the code size of the benchmarks (\ref{Apx:codesize}), case studies (\ref{Apx:examples}), and how to run \tool (\ref{Apx:running}).

\section{Conclusion and Future Work}
\label{Se:conclusion}
In this tool paper, we introduce \tool, which can fix data race errors in OpenMP programs written in C/C++ and Fortran. \tool can also remove unnecessary barriers and ordered regions in the programs while preserving correctness. We have affirmed the effectiveness of our work by attempting repair on multiple benchmark suites (consisting of 415 C/C++ and Fortran programs) using our tool. \tool was able to repair more than 80\% of the programs that had a valid data race error in them.

The technique behind \tool can support any OpenMP verifier. Leveraging other static and dynamic verifiers might help repair some of the programs that the combination of \tool and \verifier could not. More details on these verifiers have been provided in \ref{Apx:verifiers}.

\section*{Acknowledgements}
We thank the Ministry of Education, India, for their financial support.

\bibliographystyle{splncs04}
\bibliography{references.bib}

\appendix
\renewcommand{\thesection}{Appendix \Alph{section}}
\renewcommand{\thesubsection}{\Alph{section}.\arabic{subsection}}
\section{More Experiments and Results}
\label{Apx:results}

\subsection{Source Code Size}
\label{Apx:codesize}

\figref{code_lines} shows the size of the programs in terms of lines of code. The average number of lines of code for the test suite is $694.92$, and the median is $44$. $58$ programs have more than $100$ lines of code, and $184$ programs have more than $50$ lines of code.

\begin{figure}[htp]
\centering

\begin{minipage}{.45\textwidth}
    \centering
    \begin{tikzpicture}[scale=0.7]
    
    \selectcolormodel{gray}
    
    \begin{axis}[
        xbar, xmin=0, xmax=250,
        xlabel={Kernel Count},
        symbolic y coords={
            {$> 100$},
            {$81-100$},
            {$61-80$},
            {$41-60$},
            {$21-40$},
            {$<= 20$}
        },
        ytick=data,
        ylabel={Lines of Code},
        nodes near coords,
        nodes near coords align={horizontal},
        height=200pt, width=200pt
    ]
    
    \addplot coordinates {
        (58,{$> 100$})
        (5,{$81-100$})
        (67,{$61-80$})
        (87,{$41-60$})
        (177,{$21-40$})
        (21,{$<= 20$})
    };
    
    \end{axis}
    \end{tikzpicture}
    
    \caption{Lines of Code}
    \label{Fi:code_lines}
\end{minipage}
\hfil
\begin{minipage}{.5\textwidth}
    \centering
    \begin{tikzpicture}[scale=0.7]
    
    \selectcolormodel{gray}
    
    \begin{axis}[
        xbar, xmin=0, xmax=250,
        xlabel={Kernel Count},
        symbolic y coords={
            {$> 400$},
            {$301-400$},
            {$201-300$},
            {$101-200$},
            {$51-100$},
            {$<= 50$}
        },
        ytick=data,
        ylabel={LLVM IR Instructions},
        nodes near coords,
        nodes near coords align={horizontal},
        height=200pt, width=200pt
    ]
    
    \addplot coordinates {
        (78,{$> 400$})
        (27,{$301-400$})
        (38,{$201-300$})
        (88,{$101-200$})
        (123,{$51-100$})
        (51,{$<= 50$})
    };
    
    \end{axis}
    \end{tikzpicture}
    
    \caption{LLVM IR Instructions}
    \label{Fi:llvm_insts}
\end{minipage}
\vspace{-1cm}
\end{figure}
\begin{figure}[htp]
\centering

\begin{tikzpicture}[scale=0.7]

\selectcolormodel{gray}

\begin{axis}[
    xbar, xmin=0, xmax=250,
    xlabel={Kernel Count},
    symbolic y coords={
        {$> 30$},
        {$11-30$},
        {$6-10$},
        {$4-5$},
        {$1-3$},
        {$0$}
    },
    ytick=data,
    ylabel={Barrier Variables},
    nodes near coords,
    nodes near coords align={horizontal},
    height=200pt, width=200pt
]

\addplot coordinates {
    (37,{$> 30$})
    (14,{$11-30$})
    (33,{$6-10$})
    (69,{$4-5$})
    (101,{$1-3$})
    (151,{$0$})
};

\end{axis}
\end{tikzpicture}

\caption{Barrier Variables}
\label{Fi:barrier_variables}

\end{figure}
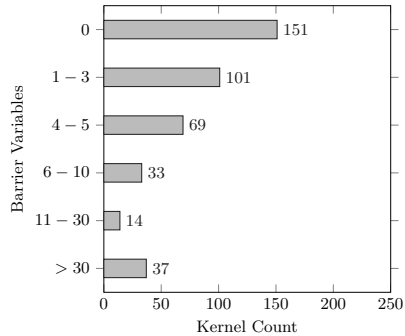

The instrumentation component of \tool works with LLVM IR and not on the source code. A line of source code could result in zero (e.g., code comments) or more LLVM IR instructions. LLVM IR was generated for $405$ programs. $10$ programs did not reach the instrumentation stage of \tool since the compiler could not generate LLVM IR for them. \figref{llvm_insts} shows the size of the programs in terms of LLVM IR instructions. The average number of LLVM IR instructions for the test suite is $1995.34$, and the median is $115$.

The repair step of \tool depends on the number of barrier variables that have been introduced during instrumentation. \figref{barrier_variables} shows the number of barrier variables introduced in the instrumentation stage of \tool. The average number of barrier variables for the test suite is $71.01$, and the median is $2$. More than $50\%$ of the programs had less than 3 barrier variables.

\subsection{Case Studies}
\label{Apx:examples}

In this section, we go over some of the programs that caused either \verifier or \tool to fail. Consider the program in \coderef{race_pointer}. Multiple threads are incrementing the value in the address pointed by the static pointer variable, thereby causing a data race. \coderef{race_llvmir} shows the simplified LLVM IR code for this. The address stored in \TT{counter} is copied to \TT{\%3}, and this variable is used to perform the increment. \tool detects that \TT{counter} is a shared variable and instruments the instructions that are using it, but does not track that the address has been copied. Because of this, line \ref{codeline:race_pointer:store} is not instrumented, and \tool is unable to fix the program.

\coderef{race_singleline} showcases an example of a shared array being read and written in the same line. Since \tool works on the LLVM IR level, such statements are split into two LLVM instructions, one for the read and another for the write. By leveraging this, \tool can fix such cases. The output message in these cases would suggest the line where the barrier needs to be inserted, but the user will have to split the line in the source program manually and insert the barrier.

\begin{minipage}{.45\textwidth}
    \begin{lstlisting}[caption=Pointer Data Race, label={Cd:race_pointer}, captionpos=b, xleftmargin=0em, escapechar=\^]
static int* counter;

int main() {
    #pragma omp parallel
        (*counter)++
}
    \end{lstlisting}
\end{minipage}
\hfil
\begin{minipage}{.45\textwidth}
    \begin{lstlisting}[caption=LLVM IR, label={Cd:race_llvmir}, captionpos=b, xleftmargin=2em, escapechar=\^]
%3 = load i32*, i32** @counter
%4 = load i32, i32* %3
%5 = add nsw i32 %4, 1
store i32 %5, i32* %3^\label{codeline:race_pointer:store}^
    \end{lstlisting}
\end{minipage}

\begin{lstlisting}[caption=OpenMP program with a data race in the same line, label={Cd:race_singleline}, captionpos=b, xleftmargin=2em, escapechar=\^]
int data[NUM_THREADS+1];
#pragma omp parallel {
    int id = omp_get_thread_num();
    data[id] = data[id+1];
}
\end{lstlisting}

\coderef{llov_fortran} and \coderef{llov_c} are two equivalent programs written in Fortran and C++, respectively. It can clearly be seen that the programs have a data race because of the read instruction at line \ref{codeline:llov_fortran:read} and the write instruction at line \ref{codeline:llov_fortran:write}.

\begin{lstlisting}[caption=OpenMP program in Fortran with a data race, label={Cd:llov_fortran}, captionpos=b, xleftmargin=2em, escapechar=\^]
program datarace
    use omp_lib

    implicit none
    integer :: id, temp, threads = 10
    integer :: data(11)

    !$omp parallel private(id,temp)
        id = omp_get_thread_num()+1
        temp = data(id+1)^\label{codeline:llov_fortran:read}^
        data(id) = temp^\label{codeline:llov_fortran:write}^
    !$omp end parallel
end program
\end{lstlisting}

\begin{lstlisting}[caption=OpenMP program in C++ with a data race, label={Cd:llov_c}, captionpos=b, xleftmargin=2em, escapechar=\^]
#include "omp.h"
#define NUM_THREADS 10

int main()
{
    int data[NUM_THREADS+1];
    #pragma omp parallel
    {
        int id = omp_get_thread_num();
        int temp = data[id+1];
        data[id] = temp;
    }
}
\end{lstlisting}

\begin{table}[h]
\caption{LLVM Instructions for the Fortran program in \coderef{llov_fortran}}
\label{Ta:llvm_fortran_insts}
\begin{center}

\def\arraystretch{1.1}
\setlength\tabcolsep{7pt}

\begin{tabular}{|l|l|}
\hline

\multicolumn{1}{|c|}{\textbf{}} & \multicolumn{1}{|c|}{\textbf{LLVM Instructions}} \\ \hline
1 & \makecell[l]{
\TT{\%0 = call i32 (...) @omp\_get\_thread\_num()} \\
\TT{\%1 = add nsw i32 \%0, 1}
} \\ \hline
2 & \makecell[l]{
\TT{\%2 = sext i32 \%1 to i64} \\
\TT{\%3 = bitcast \%struct.BSS1* @.BSS1 to i32*} \\
\TT{\%4 = getelementptr i32, i32* \%3, i64 \%2} \\
\TT{\%5 = load i32, i32* \%4, align 4}
} \\ \hline
3 & \makecell[l]{
\TT{\%6 = sext i32 \%1 to i64} \\
\TT{\%7 = bitcast \%struct.BSS1* @.BSS1 to i8*} \\
\TT{\%8 = getelementptr i8, i8* \%7, i64 -4} \\
\TT{\%9 = bitcast i8* \%8 to i32*} \\
\TT{\%10 = getelementptr i32, i32* \%9, i64 \%6} \\
\TT{store i32 \%5, i32* \%10, align 4}
} \\ \hline

\end{tabular}

\end{center}
\end{table}

Even though these programs are similar, \verifier is able to identify a data race in the C++ program but not in the Fortran program. Upon investigation, we found out that the LLVM IR generated by \app{Clang} (an \llvm frontend for  C++ programs) and \app{Flang} (an \llvm frontend for Fortran programs) are different. There are three steps in the parallel block: 

\begin{enumerate}
\item The \TT{thread\_num} is retrieved
\item The \TT{data} array is read
\item The \TT{data} array is written into
\end{enumerate}

\tableref{llvm_fortran_insts} and \tableref{llvm_c_insts} illustrate the LLVM IR instructions generated for these three steps for \coderef{llov_fortran} and \coderef{llov_c}, respectively.

\begin{table}[h]
\caption{LLVM Instructions for the C++ program in \coderef{llov_c}}
\label{Ta:llvm_c_insts}
\begin{center}

\def\arraystretch{1.1}
\setlength\tabcolsep{7pt}

\begin{tabular}{|l|l|}
\hline

\multicolumn{1}{|c|}{\textbf{}} & \multicolumn{1}{|c|}{\textbf{LLVM Instructions}} \\ \hline
1 & \makecell[l]{
\TT{\%4 = call i32 @omp\_get\_thread\_num()} \\
\TT{\%5 = add nsw i32 \%4, 1}
} \\ \hline
2 & \makecell[l]{
\TT{\%6 = sext i32 \%5 to i64} \\
\TT{\%7 = getelementptr inbounds [11 x i32], [11 x i32]* \%2, i64 0, i64 \%6} \\
\TT{\%8 = load i32, i32* \%7, align 4}
} \\ \hline
3 & \makecell[l]{
\TT{\%9 = sext i32 \%4 to i64} \\
\TT{\%10 = getelementptr inbounds [11 x i32], [11 x i32]* \%2, i64 0, i64 \%9} \\
\TT{store i32 \%8, i32* \%10, align 4}
} \\ \hline

\end{tabular}

\end{center}
\end{table}

The LLVM IR generated by \app{Clang} and \app{Flang} for the first two steps are similar. However, the third step (writing into the \TT{data} array) has a different implementation. In the LLVM IR generated for the C++ program, the code directly accesses the memory location represented by \TT{data[omp\_get\_thread\_num()]}. In contrast, the code of the LLVM IR generated for the Fortran program first computes the memory location represented by \TT{data[omp\_get\_thread\_num()+1]} and moves 4 bytes backward (since the array consists of 32-bit integers).

Cases like these reaffirm the fact that the soundness and completeness guarantees of \tool are dependent on the completeness of \verifier.

\begin{lstlisting}[caption=OpenMP program with a data race because of a write-write conflict, label={Cd:race_sameline}, captionpos=b, xleftmargin=2em, escapechar=\^]
int data[NUM_THREADS+1];
#pragma omp parallel {
    int id = omp_get_thread_num();
    data[0] = id;^\label{codeline:race_sameline:race}^
}
\end{lstlisting}

Consider the program in \coderef{race_sameline}. The data race occurs because all the threads are trying to update the first location of the array data with their corresponding thread numbers. This results in a write-write conflict on the same line number. No barrier placement can mitigate the data race occurring in this program. Even though \verifier correctly identifies the data race in these programs, \tool is unable to repair them.

\subsection{Running \tool}
\label{Apx:running}

\begin{figure}[htp]
\centering
\includegraphics[scale=0.53]{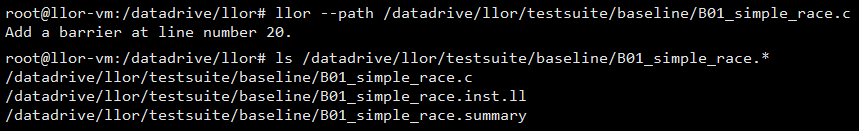}
\caption{\tool Commands}
\label{Fi:tool_commands}
\end{figure}

As described in \figref{repairer_workflow}, \tool uses multiple tools in the repair process. All these steps are abstracted from the user through a single command-line utility. After installing \tool as described in \cite{llorsrc}, the tool can be accessed from any folder. Once the program is repaired, \tool prints the needed changes in the source code and provides a fixed LLVM IR program as evidence. \figref{tool_commands} shows a sample run and changes recommended by \tool.

\begin{figure}[htp]
\centering
\includegraphics[scale=0.65]{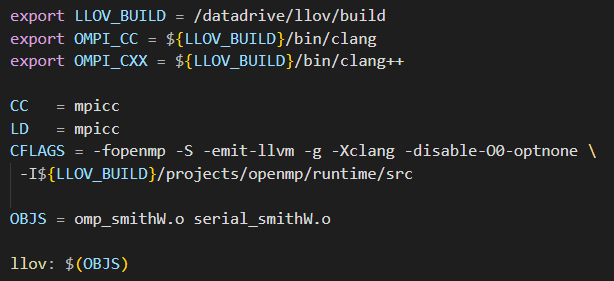}
\caption{Sample Makefile}
\label{Fi:tool_makefile}
\end{figure}

\tool also has the ability to repair programs that have multiple source files. The prerequisite for these programs is that there should be a makefile in the root folder with a target named \EM{llov}. This target should generate the LLVM IR for the various source files of the program. Examples of this makefile are provided at \cite{llorsrc}.

\begin{figure}[htp]
\centering
\includegraphics[scale=0.41]{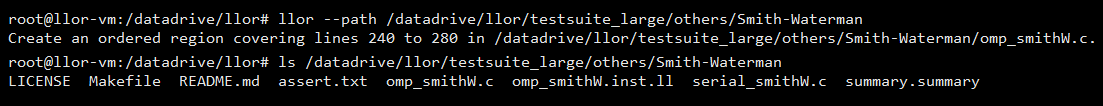}
\caption{Multifile Repair}
\label{Fi:tool_multifile}
\end{figure}

The solver type can be changed from \mhs to \maxsat through the command-line options. Additional logging can also be enabled through these options.

\begin{figure}[H]
\centering
\includegraphics[scale=0.56]{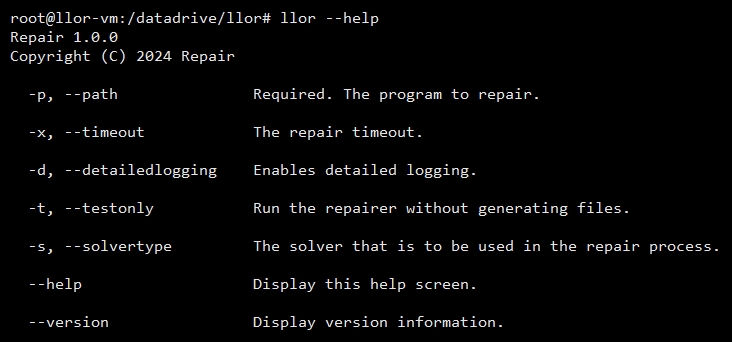}
\caption{\tool Options}
\label{Fi:tool_options}
\end{figure}

\subsection{OpenMP Verifiers}
\label{Apx:verifiers}

As mentioned in \sectref{rw_verification}, several static and dynamic tools have been proposed for identifying data races in OpenMP programs. Even though our tool uses \verifier as the verifier, in principle, the technique behind \tool works for any verifier. One possible direction that future work might take is exploring other verifiers to improve the repair coverage of \tool. In this section, we briefly introduce the other verifiers that were touched upon in \sectref{rw_verification}.

\app{ompVerify}~\cite{basupalli2011ompverify} is a polyhedral model-based static data race detection tool built on top of the \app{AlphaZ}~\cite{yuki2012alphaz} polyhedral framework. \app{DRACO}~\cite{ye2018using} is another static data race detection tool that uses the polyhedral model and is built on top of the \app{ROSE}~\cite{schordan2003source,quinlan2011rose} compiler framework. \app{PolyOMP}~\cite{chatarasi2016static,chatarasi2017extended} uses the polyhedral model to encode OpenMP loop nest information as constraints and uses the \zth~\cite{de2008z3} solver to detect race conditions. The extended version of \app{PolyOMP}~\cite{chatarasi2017extended} uses \EM{May-Happen-in-Parallel} analysis in place of \zth.

\app{SWORD}~\cite{atzeni2018sword} is a dynamic tool based on operational semantic rules and uses the OpenMP tools framework \app{OMPT}~\cite{eichenberger2013ompt}. It uses locksets to implement the semantic rules by taking advantage of the events tracked by OMPT. \app{Helgrind}~\cite{valgrind2007helgrind} and \app{Valgrind DRD}~\cite{valgrind2007drd} are dynamic race detection tools built in the \app{Valgrind}~\cite{nethercote2003valgrind} framework. Both of them use the happens-before relations to identify data races. \app{ThreadSanitizer}~\cite{serebryany2009threadsanitizer} is another dynamic data race detection tool based on \app{Valgrind}. It uses a hybrid approach of analyzing the happens-before relations and maintaining locksets for read/write operations on shared memory. \app{TSan-LLVM}~\cite{serebryany2011dynamic} is based on \app{ThreadSanitizer}. It uses \llvm to instrument the binaries instead of \app{Valgrind}. \app{Archer}~\cite{atzeni2016archer} uses both static and dynamic analysis for race detection. It uses happens-before relations, which enforces multiple runs of the program to find races. \app{Archer} reduces the analysis space for \app{TSan-LLVM} by instrumenting only parallel sections of an OpenMP program.

\end{document}